\documentclass[12pt,preprint]{aastex}
\usepackage{lscape}
\usepackage{multicol}
\newcommand{\gsim}{\raisebox{-0.3ex}{\mbox{$\stackrel{>}{_\sim} \,$}}}

\slugcomment{Version from \today}

\begin{document}

\title{High-Mass Starless Cores}

\author{T.K. Sridharan\altaffilmark{1,4}, H. Beuther\altaffilmark{1,4}, M.
Saito\altaffilmark{2,4}, F. Wyrowski\altaffilmark{3,4}, P. Schilke\altaffilmark{3,4}}
\altaffiltext{1}{Harvard-Smithsonian Center for Astrophysics, 60 Garden Street, MS 78, Cambridge, MA 02138, USA.}
\altaffiltext{2}{National Astronomical Observatory of Japan, 2-21-1 Osawa, Mitaka, Tokyo, 181-8588, Japan}
\altaffiltext{3}{Max-Planck-Institut f\"ur Radioastronomie, Auf dem H\"ugel 69, 53121 Bonn, Germany}
\altaffiltext{4}{emails: tksridha@cfa.harvard.edu, hbuether@cfa.harvard.edu, Masao.Saito@nao.ac.jp, schilke@mpifr-bonn.mpg.de, wyrowski@mpifr-bonn.mpg.de} 
\begin{abstract}

We report the identification of a sample of potential High-Mass Starless Cores (HMSCs).
The cores were discovered by comparing images of the fields containing candidate 
High-Mass Protostellar Objects (HMPOs) at 1.2mm and mid-infrared (8.3$\mu$m; MIR) 
wavelengths.  While the HMPOs are detected at both wavelengths, several cores emitting 
at 1.2mm in the same fields show absorption or no emission at the MIR wavelength. 
We argue that the absorption is caused by cold dust. The estimated masses of 
a few 10$^2M_\odot$ - 10$^3M_\odot$ and the lack of IR emission suggests 
that they may be massive cold cores in a pre-stellar phase, which could presumably 
form massive stars eventually.  Ammonia (1,1) and (2,2) observations of the cores 
indicate smaller velocity dispersions and lower rotation temperatures compared 
to HMPOs and UCH{\sc ii} regions suggesting a quiescent pre-stellar stage. We propose 
that these newly discovered cores are good candidates for the HMSC 
stage in high-mass star-formation. This sample of cores will allow us to study 
the high-mass star and cluster formation processes at the earliest evolutionary
stages.

\end{abstract}


\section {Introduction}

The area of high-mass star-formation is witnessing significant progress.
Systematic studies have uncovered several High-Mass Proto-Stellar 
objects (HMPOs) in a pre-Ultra Compact
H{\sc ii} (UCH{\sc ii}) region phase (Molinari et al 1996,2002; Sridharan et al 2002;
Beuther et al 2002a). The ubiquity of the outflows in these objects 
points to essential similarities between the high- and the low-mass star-formation 
processes viz., the presence of a disk-accretion phase (Zhang et al
2005; Beuther et al 2002b). However, the
characteristics of the High-Mass Starless Core (HMSC) stage which must 
precede the HMPO stage has barely been studied (Evans et al 2002). In contrast, recent 
studies of low-mass starless cores have begun to constrain the initial 
conditions for low-mass star-formation (Motte et al 1998; Bacmann et al 2000; 
Alves, Lada \& Lada, 2001; Harvey et al, 2003). Similar studies of HMSCs are 
important in answering one of the central questions in
star-formation: how do star-formation processes produce the stellar 
Initial Mass Function? Since high-mass stars form in 
clusters, and a major fraction of all stars form in such environments,
the structure of the HMSCs represents the initial conditions relevant for
star-formation in general.

In this letter, we present and study a sample of potential HMSCs identified 
by combining images at millimetre and mid-infrared wavelengths of HMPO
fields.  The candidate 
HMSCs belong to the general class of infrared-dark clouds (IRDCs), studied recently by others (Perault et al 1996; Egan et al 1998; 
Carey et al 2000, Bacmann et al 2000; Teyssier, Hennebelle \& Perault 2002, 
Garay et al 2004). At the outset, we want to clarify two points: we use 
the terms {\it starless} and {\it pre-stellar} to indicate the lack of massive 
star-formation ($\gsim$ 8 M$_{\odot}$; spectral types B2 or earlier); the 
objects discussed here are {\it candidates} for the HMSC stage although this is
not mentioned every time for brevity.

\section {Source Sample}
The data used to identify the HMSCs consist of continuum images at 
1.2mm and 8.3$\mu$m wavelengths, of a sample of 69 HMPOs we have been studying (Sridharan 
et al 2002, Beuther et al 2002a,b, Williams, Fuller \& Sridharan, 2004, 2005). 
The 1.2mm images obtained with the MAMBO array at the IRAM 30M telescope have been
 described in detail by Beuther et al (2002a). 
The 8.3$\mu$m images are the archival data from the Mid-Course Space Experiment 
(MSX; Egan et al 1998),  available at the MSX web site 
(http://www.ipac.caltech.edu/ipac/msx/msx.html).

The 1.2mm images, each covering a $\sim$ 5$^\prime\times$ 5$^\prime$ field
around an HMPO, detected multiple sources in a number of fields with an 
average multiplicity rate of 2.2 (Beuther et al 2002). We argued in 
Sridharan et al (2002) and Beuther et al (2002a) that the objects in our 
HMPO sample are likely to be in various stages of evolution. This is expected, 
given the clustered mode in which high-mass star-formation is thought to occur. 
Further, it is conceivable that an earlier starless stage in the high-mass 
star-formation process should occur in the same fields.  With this motivation, 
we compared  the 1.2mm and the
MSX A band (8.3$\mu$m) images of the HMPO fields.  Strong 
 emission at the 
two wavelengths was found to be typical of the HMPOs, within reasonable 
position uncertainties. Significant position offsets between MSX and mm 
emission are accounted for in terms of a cluster setting and varying
evolutionary stages (Sridharan et al 2002). However, in many fields, we also 
discovered 1.2mm emission features associated with either MSX absorption
or no detectable emission. Figure 1 shows three examples. The 
strongest mm emission near
the center of the field in Figure 1a is associated
with the HMPO-IRAS18385-0512 which is also detected at 8.3$\mu$m. 
The mm emission feature seen to the south east of the HMPO 
is associated with absorption at 8.3$\mu$m. 
Figures 1b (HMPO-IRAS18223-1243; also studied by Garay et al 2004)
  and 1c present more examples, to indicate the range of morphologies,
showing filamentary structures and a field of multiple cores where a stellar 
cluster may be forming.  Following similar observations
of low-mass starless cores (for example, Bacmann et al, 2000), we suggest 
that the absorption is caused by the 
same cold dust emitting at 1.2mm. The similarity of the
morphologies of the emission and absorption 
supports this hypothesis.  In the case of
IRDC-18385-0512-3, shown in Figure 1a, using a temperature of 15 K 
(see section 3.1), and the 1.2mm flux,  a mass of 344 M$_{\odot}$ is
estimated, making it a potential High-Mass Starless Core.


We have identified a total of 56 candidate HMSCs,
listed in Table 1 along with kinematic distances, 1.2mm fluxes and mass estimates. 
Each object is labelled with the HMPO field and the mm-emission clump number
from Beuther et al (2002a). For completeness, we have included 
26 prominent MSX absorption objects either lying outside the 
1.2mm images or where the images are too noisy, extending the numbering. 
For mass estimates (30 objects), 
as for the HMPOs (Beuther et al 2002a, 2005; Hildebrand 1983), we used a 
value of 2 for the spectral 
index of dust emissivity $\beta$, 100 for the gas to dust ratio, 
0.1 $\mu$m for grain size and 3 g cm$^{-3}$ for grain mass density ($\kappa$ = 0.4 cm$^2$/gm of dust), in
addition to a temperature of 15K (section 3.1).  In case of distance
ambiguity, the near distance has been used - at the far distance MIR absorption is unlikely. 
As seen in the table, the cores are massive, with masses 
in the range of a few 10$^2M_\odot$ - 10$^3M_\odot$.

An examination of the High Resolution processed IRAS images (HIRES; Aumann, 
Fowler, Melnyk 1990) of the fields failed to detect any emission from dust 
heated by star-formation activity.  Based on the Arcetri Catalog of water 
masers (Valdettaro et al 2001), the General Catalog of 
6.7GHz methanol masers (Pestalozzi, M.R., Minier V. and Booth, R.S., 2005) and 
interferometric 
maser observations of the HMPO fields (Beuther et al 2002c), we find two
cases -  
IRAS18102-1800-1 \& IRAS18151-1208-2 - to be associated  with 
methanol/water maser emission respectively, implying 
star-formation in these cores. This suggests that the cores 
presented here contain objects in multiple evolutionary stages. However, all these 
stages represent phases before the pre-UCH{\sc ii}/HMPO phase, containing
either low-mass young stellar objects or early pre-cursors to the HMPOs. A further 
implication is that the maser activity may turn on very early in the star-formation 
process, and a majority of the cores presented here may be in a younger phase.

\section {Observations \& Results}

To characterize the state of the dense gas in the sample, 
a subset of 34 objects from Table 1 was observed with the Efflesberg 100-m 
telescope for NH$_3$ emission in Dec 2002.  The (1,1) and (2,2)
inversion lines were observed simultaneously in frequency switched 
mode with a velocity resolution of $\Delta v=0.25$\,kms$^{-1}$ and $\sim$ 4 minutes
of integration time per position. The system temperatures were
mostly below 50\,K.  Both lines were detected in all the observed
sources.  For 18 sources with good quality spectra,
rotation temperatures were derived following Ungerechts et al. (1986). 
For temperatures below 20\,K, as here, kinetic temperatures are only 
marginally higher (Danby et al., 1988).  Figure 2 shows an example NH$_3$(1,1) 
spectrum, for the object IRAS18385-0512-3 (whose image is shown in Figure 1a). 
This core has a small linewidth of 1.1 kms$^{-1}$.


\subsection{Ammonia line-widths and temperatures}

Figure 3 compares the distributions of the ammonia line widths and
rotation temperatures for the HMSCs,  HMPOs and
UCH{\sc ii} regions (Sridharan et al 2002, Churchwell et al, 1990).  
The HMSCs exhibit a smaller line width with a median of 
1.5 kms$^{-1}$ (mean 1.6 kms$^{-1}$) compared to 1.9 kms$^{-1}$ (mean
2.1kms$^{-1}$) for the HMPOs. This suggests smaller
internal motions and therefore more quiescent conditions, supporting the pre-stellar nature 
of the HMSCs. The rotation temperatures indicate
colder conditions in HMSCs with a median value of 16.9 K (mean 15.3) compared to 18.5 K (mean 22.5) 
for HMPOs, again suggesting more quiescent conditions in the HMSCs. 
The corresponding ammonia line widths and temperatures for UCH{\sc ii} regions are 
3 kms$^{-1}$ and 22 K (Churchwell et al 1990). Thsese results support the picture where 
the objects evolve from HMSCs to HMPOs to UC-H{\sc ii} regions.  

Although the ammonia rotation temperatures for the HMSCs are only marginally lower than 
for the HMPOs, where grey-body dust temperatures were estimated to be much higher 
(Sridharan et al, 2002), we believe that the dust in the HMSCs
is cooler, and well characterized by the gas temperature.  
For the HMPOs, the best fits to the spatial 
and the spectral energy distributions at FIR and mm- wavelengths 
imply temperatures increasing inwards (Williams, Fuller \& Sridharan,
2005). Ammonia emission traces the cooler outer regions of the cores, leading to lower
rotation temperature estimates. 
In contrast, for externally heated cores, as suggested here for the HMSCs, the 
dust temperature is expected to be similar to the gas temperature in the
outer regions or may even 
be marginally lower (Li, Goldsmith \& Menten, 2003). 
The lack of FIR emission towards the HMSCs - in cases 
where the the dominant HMPO in the field is not nearby, no 
emission is seen even at 60$\mu$m or 100$\mu$m 
wavelengths in the HIRES images - indicates the absence of dust heated by an embedded object. 

The distributions of ammonia rotation temperatures and linewidths of the HMSCs and HMPOs
were subjected to a Kolmogorov-Smirnov test, resulting in 0.90 and 0.62 for the probabilities
that they are different.
The masses of HMSCs are distributed over a similar range as the HMPOs. 
Virial masses estimated from ammonia line widths and
1.2mm sizes are lower, but comparable to dust derived masses, consistent
with the HMSCs being in equilibrium. 
Because of the many
uncertainties in mass estimates in both the cases, we do not make further 
comparisons. 

As already noted, we use the term {\it starless} here to imply the absence of a massive 
star. 
From the IRAS flux limits, a luminosity upper limit of $\sim$100 L$_{\odot}$ is
derived for a distance of 4 kpc, corresponding to early A spectral
types ($\sim$ 2-3 M$_{\odot}$).  Assuming a star-formation efficiency of 30\%, 
an HMSC mass of 550 M$_{\odot}$ (sample median), and an IMF
with power-law indices 2.3 and 1.3 for masses $>$ 0.5 M$_{\odot}$ and
0.08 - 0.5 M$_{\odot}$ respectively (Kroupa, 2004), we estimate that one star of mass
$\gsim$ 20 M$_{\odot}$ could form in an HMSC. While there is no current 
evidence for massive star-formation in these cores from their infrared fluxes, more 
sensitive seraches or studies like the Spitzer GLIMPSE survey will presumably uncover 
lower mass star-formation or hitherto unknown early stages of high-mass
star-formation.

In summary, we have identified a sample of candidate high-mass starless cores, 
likely to contain objects in multiple stages of evolution prior to the formation of 
massive stars. The sample provides an opportunity to determine initial conditions for 
high-mass star/cluster formation and star-formation in general. The study of their internal structure at multiple 
wavelengths using instruments like the Sub-millimeter Array (SMA) and the Spitzer 
telescope will pave the way for the understanding 
the origin of the stellar Initial Mass Function (IMF), a central puzzle in star-formation. 
\acknowledgements
\noindent H.B. thanks the Emmy-Noether-Program of the Deutsche
Forschungsgemeinschaft for financial support (DFG, grant BE2578/1).


\newpage

\includegraphics[angle=-90, scale=0.8]{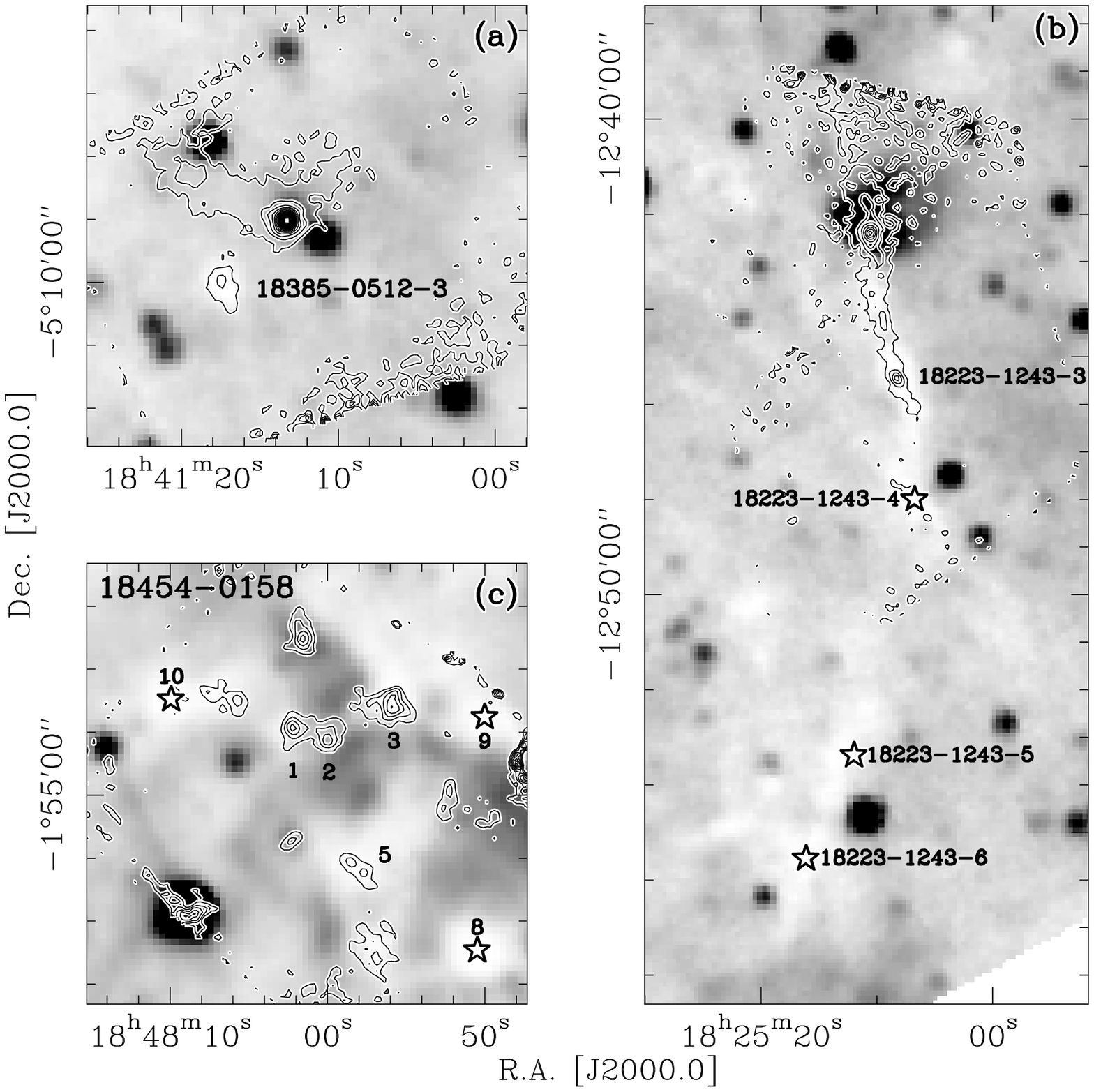}
\figcaption{\small MSX A-band (8$\mu$m) images (black is bright) with 1.2mm 
emission contours: (a) IRDC-18385-0512-3; (b) and (c) show fields 18223-1243 and
18454-0158 with filaments and multiple cores. The star symbols mark cores lacking 
good 1.2mm measurements.}

\newpage

\includegraphics[angle=-90, scale =0.65]{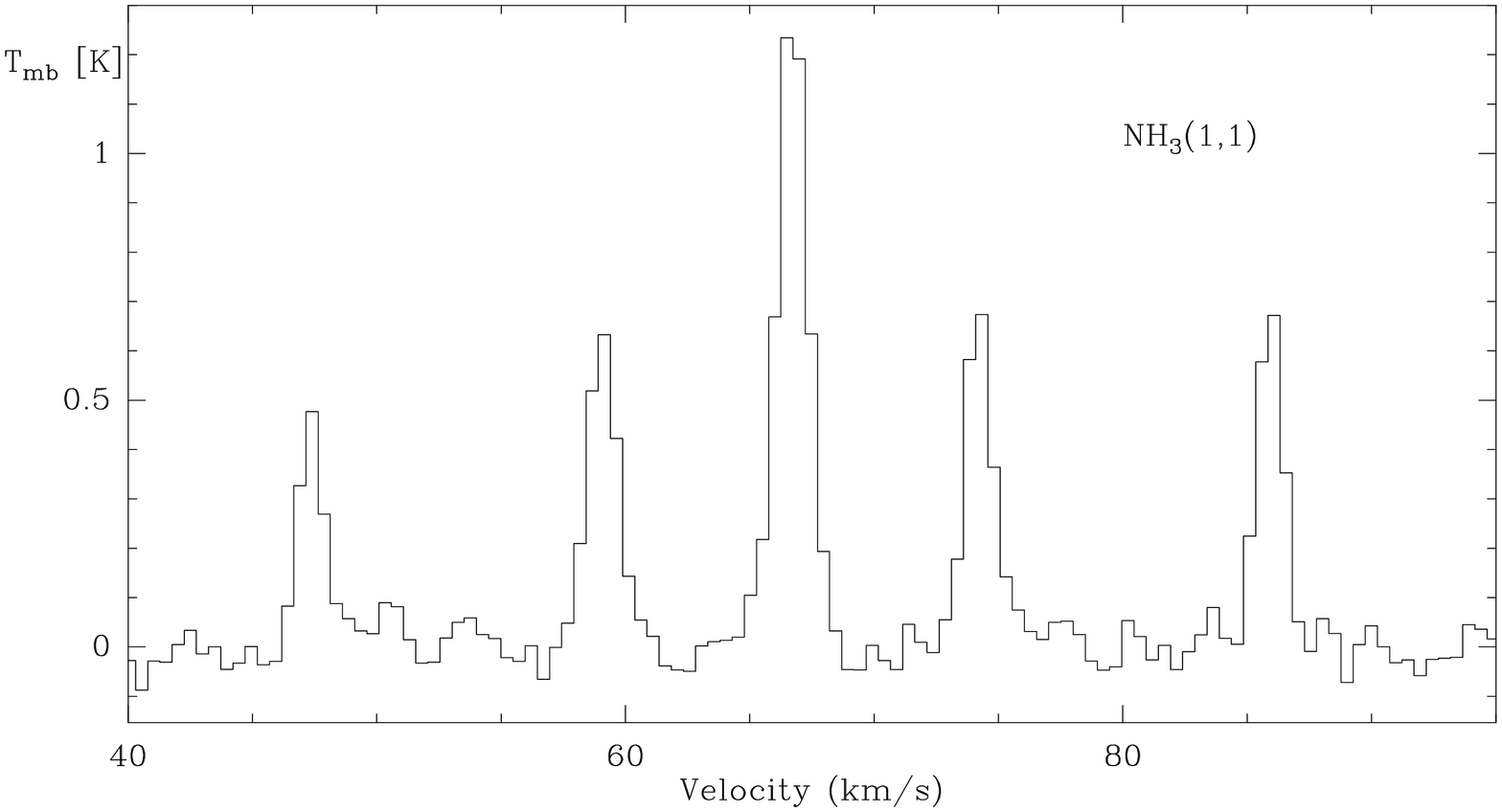}
\figcaption{\small  An NH$_3$(1,1) spectrum for IRAS18385-0512-3.}

\newpage

\includegraphics[angle=-90, scale = 0.75]{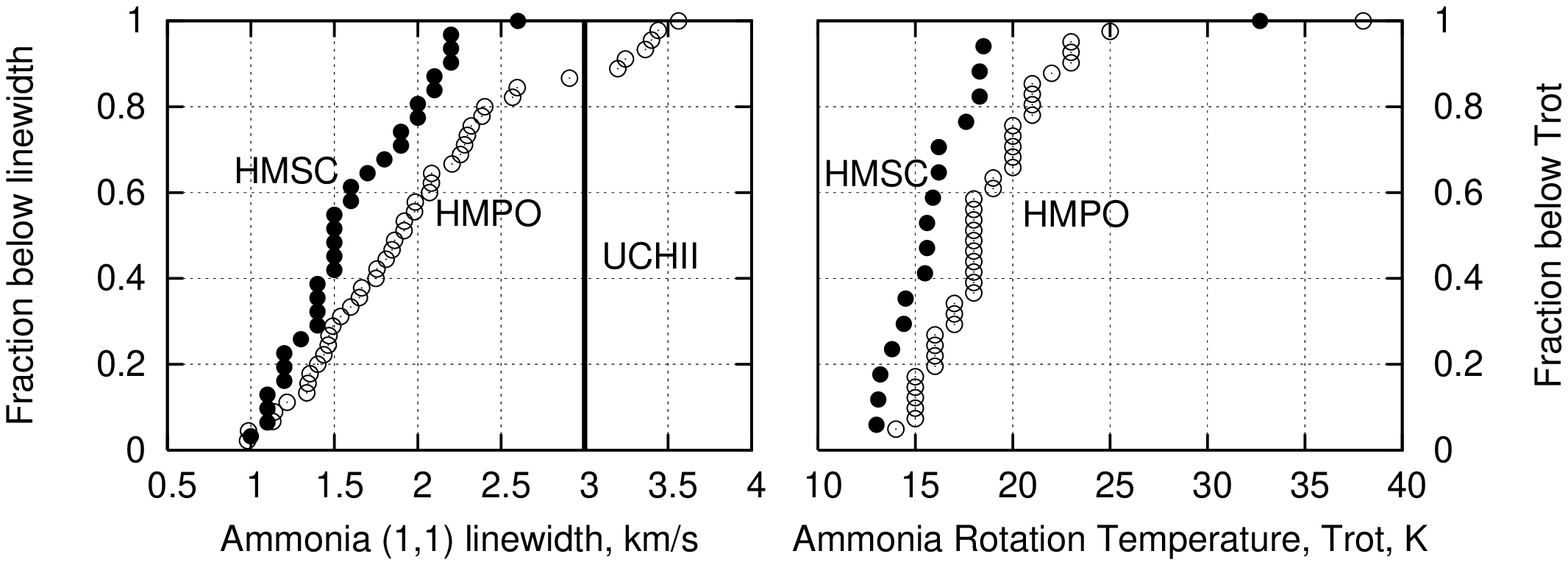}
\figcaption{\small  The cumulative distributions of NH$_3$(1,1) line widhts 
for HMSCs and HMPOs with the median for UCH{\sc ii} regions shown as a vertical
line ({\it left}). The cumulative distributions of NH$_3$ rotation
temperatures for HMSCs and HMPOs ({\it right}).}
\newpage

\begin{deluxetable}{lllllllllr}
\tablecolumns{10}
\tablewidth{0pc}
\tablecaption{Table 1 -- Candidate High-Mass Starless Cores}
\tabletypesize{\footnotesize}
\tablehead{
\colhead{No}&\colhead{Name - Field$-$object\#}         &\colhead{ra}         &\colhead{dec}     &\colhead{D}
&\colhead{S$_{1.3}$}   &\colhead{$v_{lsr}$}  &\colhead{$\Delta v$}   &\colhead{T} &\colhead{M$_{15K}$} \\
\colhead{}   &\colhead{}                         &\multicolumn{2}{c}{J2000.0}  &\colhead{kpc}
&\colhead{Jy}  &\colhead{km/s}  &\colhead{km/s} &\colhead{K} &\colhead{M$_{\odot}$}} 
\startdata
01 &IRDC-IRAS18089$-$1732-3 &18 11 45.3 &$-$17 30 38 &3.6 &0.3 &34.1 &2.2 &15.6  & 241  \\
02 &IRDC-IRAS18090$-$1832-2 &18 12 02.0 &$-$18 31 27 &6.6 &0.2 &      &     & & 541 \\
03 &IRDC-IRAS18102$-$1800-1 &18 13 11.0 &$-$18 00 23 &2.6 &3.3 &14.0  &     & & 1385 \\
04 &IRDC-IRAS18151$-$1208-2 &18 17 50.3 &$-$12 07 54 &3.0 &2.6 &29.3 &2.2 &18.5  & 1453 \\
05 &IRDC-IRAS18182$-$1433-2 &18 21 14.9 &$-$14 33 06 &3.6\tablenotemark{a} &0.3 &40.5  &1.4 &15.5  & 241 \\
06 &IRDC-IRAS18182$-$1433-3 &18 21 17.5 &$-$14 29 43 &    &    &59.4  &1.2    &  \\
07 &IRDC-IRAS18182$-$1433-4 &18 21 14.0 &$-$14 34 20 &    &    &      &     & \\
08 &IRDC-IRAS18223$-$1243-2 &18 25 10.0 &$-$12 44 00 &3.7 &0.6 &      &     & & 510 \\
09 &IRDC-IRAS18223$-$1243-3 &18 25 08.3 &$-$12 45 27 &3.7 &0.8 &45.3 &2.2 &32.7  & 245\tablenotemark{b} \\
10 &IRDC-IRAS18223$-$1243-4 &18 25 06.8 &$-$12 48 00 &3.7 &0.3 &45.4 &1.4 &13.0  & 255 \\
11 &IRDC-IRAS18223$-$1243-5 &18 25 12.0 &$-$12 53 23 &    &    &44.7&1.1    &  \\
12 &IRDC-IRAS18223$-$1243-6 &18 25 16.2 &$-$12 55 33 &    &    &45.3/50.5&1.4/2.1    &  \\
13 &IRDC-IRAS18247$-$1147-3 &18 27 31.0 &$-$11 44 46 &6.7 &0.4 &      &     &  & 1115 \\
14 &IRDC-IRAS18264$-$1152-2 &18 29 21.0 &$-$11 51 55 &    &    &      &     &  \\
15 &IRDC-IRAS18264$-$1152-3 &18 29 27.0 &$-$11 50 58 &    &    &      &     &  \\
16 &IRDC-IRAS18306$-$0835-3 &18 33 32.1 &$-$08 32 28 &2.5\tablenotemark{a} &0.8&31.6 &1.6 &13.1  & 310 \\
17 &IRDC-IRAS18306$-$0835-4 &18 33 34.8 &$-$08 31 20 &     &   &32.1&2.0    &    \\
18 &IRDC-IRAS18308$-$0841-2 &18 33 34.3 &$-$08 38 42 &4.9 &0.6 &      &     &  & 895 \\
19 &IRDC-IRAS18308$-$0841-3 &18 33 29.3 &$-$08 38 17 &4.9 &0.7 &73.7 &1.9 &16.2  & 1044 \\
20 &IRDC-IRAS18308$-$0841-5 &18 33 34.4 &$-$08 37 36 &4.9 &0.2 &76.7 &1.5 &14.5  & 298 \\
21 &IRDC-IRAS18308$-$0841-6 &18 33 34.9 &$-$08 36 04 &     &   &76.9&1.6    &   \\
22 &IRDC-IRAS18310$-$0825-4 &18 33 39.5 &$-$08 21 10 &5.2 &0.5 &86.5 &1.7 &18.3  & 840 \\
23 &IRDC-IRAS18337$-$0743-3 &18 36 18.2 &$-$07 41 00 &4.0 &0.3 &55.6 &1.8 &15.6  & 298 \\
24 &IRDC-IRAS18337$-$0743-4 &18 36 29.9 &$-$07 42 05 &    &    &55.2  &2.1 &  \\
25 &IRDC-IRAS18337$-$0743-5 &18 36 41.0 &$-$07 39 56 &    &    &      &     &  \\
26 &IRDC-IRAS18337$-$0743-6 &18 36 36.0 &$-$07 42 17 &    &    &      &     &  \\
27 &IRDC-IRAS18337$-$0743-7 &18 36 19.0 &$-$07 41 48 &    &    &      &     & \\
28 &IRDC-IRAS18348$-$0616-8 &18 37 14.7 &$-$06 17 25 &    &    &109.2&     & \\
29 &IRDC-IRAS18348$-$0616-2\tablenotemark{c} &18 37 27.6 &$-$06 14 08 &6.3 &0.5 &      &     & & 1232 \\
30 &IRDC-IRAS18385$-$0512-3 &18 41 17.4 &$-$05 10 03 &4.3\tablenotemark{a} &0.3 &66.7 &1.1 &14.4  & 344 \\
31 &IRDC-IRAS18431$-$0312-3 &18 45 45.0 &$-$03 08 56 &    &    &      &     & \\
32 &IRDC-IRAS18431$-$0312-4 &18 45 53.0 &$-$03 09 01 &    &    &      &     & \\
33 &IRDC-IRAS18437$-$0216-3 &18 46 21.9 &$-$02 12 24 &6.6\tablenotemark{d} &0.3 &110.2/96.4 &1.4/1.24 &15.9/13.8  &  811 \\
34 &IRDC-IRAS18437$-$0216-7 &18 46 22.0 &$-$02 14 10 &    &    &      &     &  \\
35 &IRDC-IRAS18440$-$0148-2 &18 46 31.0 &$-$01 47 08 &    &    &      &     &  \\
36 &IRDC-IRAS18445$-$0222-4 &18 47 14.6 &$-$02 15 44 &    &    &88.2 &1.0     &  \\
37 &IRDC-IRAS18447$-$0229-3 &18 47 42.0 &$-$02 25 12 &    &    &101.0 & 1.1    &  \\
38 &IRDC-IRAS18447$-$0229-4 &18 47 38.9 &$-$02 28 00 &    &    &98.6/111.8&1.3     &   & \\
39 &IRDC-IRAS18447$-$0229-5 &18 47 31.4 &$-$02 26 46 &    &    &104.0 &1.2     &  \\
40 &IRDC-IRAS18454$-$0158-1 &18 48 02.1 &$-$01 53 56 &4.7\tablenotemark{d} &0.4 &51.9/99.6 &1.5/1.9     &  & 549 \\
41 &IRDC-IRAS18454$-$0158-3 &18 47 55.8 &$-$01 53 34 &6.0\tablenotemark{a} &0.7 &93.7/97.6 &     &  & 1565 \\
42 &IRDC-IRAS18454$-$0158-5 &18 47 58.1 &$-$01 56 10 &6.0\tablenotemark{a} &0.3 &93.9 &1.5 &17.6  & 671 \\
43 &IRDC-IRAS18454$-$0158-8 &18 47 50.5 &$-$01 57 28 &    &    &94.6 &     &  \\
44 &IRDC-IRAS18454$-$0158-9 &18 47 50.0 &$-$01 53 46 &    &    &95.6/101.8 &     &  \\
45 &IRDC-IRAS18454$-$0158-10&18 48 10.0 &$-$01 53 29 &    &    &100.4 &     &  \\
46 &IRDC-IRAS18454$-$0158-11&18 48 07.0 &$-$01 53 25 &    &    &      &     &  \\
47 &IRDC-IRAS18454$-$0158-12&18 48 05.7 &$-$01 53 28 &    &    &      &     &  \\
48 &IRDC-IRAS18460$-$0307-3 &18 48 36.0 &$-$03 03 49 &5.2 &0.3 &      &     &  & 504 \\
49 &IRDC-IRAS18460$-$0307-4 &18 48 46.0 &$-$03 04 05 &5.2 &0.3 &      &     & & 504 \\
50 &IRDC-IRAS18460$-$0307-5 &18 48 47.0 &$-$03 01 29 &5.2 &0.5 &      &     & & 840 \\
51 &IRDC-IRAS18530$+$0215-2 &18 55 29.0 &$+$02 17 43 &    &    &      &     &  \\
52 &IRDC-IRAS19175$+$1357-3\tablenotemark{e} &19 19 52.1 &$+$14 01 52 &1.1\tablenotemark{a} &0.2 &7.7 &1.5 &15.6  & 15 \\
53 &IRDC-IRAS19175$+$1357-4\tablenotemark{e} &19 19 50.6 &$+$14 01 22 &1.1\tablenotemark{a} &0.2 &7.7 &1.5 &13.2  & 15 \\
54 &IRDC-IRAS19410$+$2336-2 &19 43 10.2 &$+$23 45 04 &2.1 &1.4 &21.4 &2.0 &18.3  & 383 \\
55 &IRDC-IRAS20081$+$2720-1\tablenotemark{e} &20 10 13.0 &$+$27 28 18 &0.7 &0.6 & 5.5 &     & & 18 \\
56 &IRDC-IRAS22570$+$5912-3 &22 58 55.1 &$+$59 28 33 &5.1 &0.5 &$-$47.1 &2.6 &16.2  &  807 \\
\enddata
\tablenotetext{a}{different velocities, kinematic distances and masses compared to those in Sridharan et
al (2002) and Buether et al (2002) }
\tablenotetext{b}{a dust temperature of 32.7 K was used to compute this mass; significantly higher temperature than the rest of the sample}
\tablenotetext{c}{the RA reported in Beuther et al (2002) for this object is in error by $\sim$30$"$}
\tablenotetext{d}{an average of the kinematic distances for the two velocity components}
\tablenotetext{e}{may not qualify as high-mass; included for completeness}
\end{deluxetable}


\newpage 
\begin{thebibliography}{}
\bibitem[Alves et al 2001]{Alves 2001}Alves, J. F., Lada, C. J., Lada, E. A., 2001, Nature, 409, 159
\bibitem[Aumann et al.(1990)]{1990AJ.....99.1674A} Aumann, H.~H., Fowler, 
J.~W., \& Melnyk, M.\ 1990, \aj, 99, 1674
\bibitem[Bacmann et al 2000]{Bacmann 2000} Bacmann, A., Andre, P., Puget, J.-L., Abergel, A., 
Bontemps, S., Ward-Thompson, D., 2000, A\&A, 361, 555
\bibitem[Buether et al 2002a]{Beuther 2002a} Beuther H., Schilke P., Menten K.M ., Motte, F., Sridharan T.K., Wyrowski F., 2002a, ApJ,566,945.
\bibitem[Buether et al 2005]{Beuther 2005} Beuther H., Schilke P., Menten K.M ., Motte, F., Sridharan T.K., Wyrowski F.,
2005, ApJ, in press (Erratum).
\bibitem[Beuther et al 2002b]{Beuther 2002b}Beuther, H., Schilke, P., Sridharan, T. K., Menten, K. M., 
Walmsley, C. M., Wyrowski, F., 2002, A\&A, 383, 892.
\bibitem[Beuther et al.(2002c)]{2002A&A...390..289B} Beuther, H., Walsh, A., 
Schilke, P., Sridharan, T.~K., Menten, K.~M., \& Wyrowski, F.\ 2002c, \aap, 
390, 289 
\bibitem[Carey et al 2000]{Carey 2000}Carey, S., Feldman, R., Redman, R., Egan, M.,
 Macleod, J., Price, S., 2000, ApJ, L157
\bibitem[Churchwell et al.(1990)]{1990A&AS...83..119C} Churchwell, E., 
Walmsley, C.~M., \& Cesaroni, R.\ 1990, \aaps, 83, 119 
\bibitem[Danby et al.(1988)]{1988MNRAS.235..229D} Danby, G., Flower, D.~R., 
Valiron, P., Schilke, P., \& Walmsley, C.~M.\ 1988, \mnras, 235, 229 
\bibitem[Egan et al.(1998)]{1998ApJ...494L.199E} Egan, M.~P., Shipman, 
R.~F., Price, S.~D., Carey, S.~J., Clark, F.~O., \& Cohen, M.\ 1998, \apjl, 
494, L199 
\bibitem[Evans et al.(2002)]{2002ASPC..267...17E} Evans, N.~J., Shirley, 
Y.~L., Mueller, K.~E., \& Knez, C.\ 2002, ASP Conf.~Ser.~267: Hot Star 
Workshop III: The Earliest Phases of Massive Star Birth, 267, 17 
\bibitem[Harvey et al 2003]{Harvey 2003}Harvey, D. W. A., Wilner, D. J., Lada, C. J., Myers, P. C., 
Alves, J. F., 2003, ApJ, 598, 1112
\bibitem[Garay et al 2004]{Garay 2004} Garay, G., Faundez, S., Mardones, D. et al 2004, ApJ, 610, 313.
\bibitem[Kroupa(2004)]{2004NewAR..48...47K} Kroupa, P.\ 2004, New Astronomy 
Review, 48, 47
\bibitem[Li et al.(2003)]{2003ApJ...587..262L} Li, D., Goldsmith, P.~F., \& 
Menten, K.\ 2003, \apj, 587, 262 
\bibitem[Molinari et al 1996]{Molinari 1996} Molinari, S., Brand, J., Cesaroni, R., Palla, F., 1996, 
A\&A, 308, 573.
\bibitem[Molinari et al 2002]{Molinari 2002} Molinari, S, Testi, L, Rodriguez, L. F., Zhang, Q., 
2002, ApJ, 570, 758.
\bibitem[Motte et al 1998]{Motte 1998} Motte, F., Andre, P., Neri, R., 1998, A\&A, 336, 150.
\bibitem[Pestalozzi et al.(2005)]{2005A&A...432..737P} Pestalozzi, M.~R., 
Minier, V., \& Booth, R.~S.\ 2005, \aap, 432, 737 
\bibitem[Perault et al 1996]{Perault 1996} Perault, M., Omont, A., Simon, G., 1996, A\&A, 315, l165
\bibitem[Sridharan et al 2002]{Sridharan 2002} Sridharan, T. K., Beuther, H., Schilke, P., 
Menten, K. M., Wyrowski, F., 2002, ApJ, 566, 931.
\bibitem[Williams et al.(2005)]{2005A&A...434..257W} Williams, S.~J., 
Fuller, G.~A., \& Sridharan, T.~K.\ 2005, \aap, 434, 257 
\bibitem[Williams et al.(2004)]{2004A&A...417..115W} Williams, S.~J., 
Fuller, G.~A., \& Sridharan, T.~K.\ 2004, \aap, 417, 115 
\bibitem[Teyssier et al  2002]{teyssier 2002}Teyssier, D., Hennebelle, P., Perault, M., 2002, 382, 624
\bibitem[Ungerechts et al.(1986)]{1986A&A...157..207U} Ungerechts, H., 
Winnewisser, G., \& Walmsley, C.~M.\ 1986, \aap, 157, 207 
\bibitem[Zhang et al.(2005)]{2005ApJ...625..864Z} Zhang, Q., Hunter, T.~R., 
Brand, J., Sridharan, T.~K., Cesaroni, R., Molinari, S., Wang, J., \& 
Kramer, M.\ 2005, \apj, 625, 864 
\end{thebibliography}
\end{document}